\documentclass[11pt, a4paper, reqno]{article}

\usepackage[english]{babel}										
\usepackage{etoolbox} 
\usepackage{a4wide}
\usepackage[utf8]{inputenc}										
\usepackage[T1]{fontenc}										
\usepackage{todonotes}
\usepackage{geometry}
\usepackage{caption}
\usepackage{hyperref}

\usepackage{booktabs,multicol,multirow,tabularx,array}  
\usepackage{amsmath,amsfonts,amssymb,amsthm,cancel,siunitx,
calculator,calc,mathtools,empheq,latexsym}

\usepackage{epsfig,tikz,float}		            
\usepackage[font=small,labelfont=bf]{caption} 
\usepackage{subcaption} 

\usepackage{natbib}
\usepackage{setspace}
\def\spacingset#1{\renewcommand{\baselinestretch}%
{#1}\small\normalsize} 
\spacingset{2}

\patchcmd{\subsection}{-.5em}{.01\baselineskip}{}{}
\patchcmd{\subsubsection}{-.5em}{.01\baselineskip}{}{}
\makeatother
\title{\renewcommand{\baselinestretch}{1.17}\normalsize\bf%
\uppercase{Addressing Spatial Confounding in geostatistical regression models: An R-INLA approach}
}
\author{%
Jérémy Lamouroux$^{1*}$, Alizée Geffroy$^{1}$, Sébastien Leblond$^{2}$, Caroline Meyer$^{2}$ \ and \ Isabelle Albert$^{1}$
}

\begin{document}
\date{}
\maketitle
\vspace{-0.6cm}
$^1$Université Paris-Saclay, AgroParisTech, INRAE, UMR MIA P-S, 91120 Palaiseau, France\

$^2$PatriNat (OFB-MNHN), 12 rue Buffon, 75005 Paris, France\

* Author to whom correspondence should be addressed.\

E-mail: jeremy.lamouroux@inrae.fr\\

\noindent
{\small{\bf ABSTRACT}
\begin{enumerate}
    \item[1.] Spatial confounding is a phenomenon that has been studied extensively in recent years in the statistical literature to describe and mitigate apparent inconsistencies between the results obtained by regression models with and without random spatial effects. While the most common solutions target almost exclusively areal data or geostatistical data modelling by splines, we aim to extend some resolution methods in the context of geostatistical data modelling by Gaussian Markov Random Fields (GMRF) using R-INLA methodology.
    
    \item[2.] First, we present three approaches for alleviating spatial confounding: Restricted Spatial Regression (RSR), Spatial+, and its recent simplified version, called here Spatial+ 2.0. We show how each can be implemented from geostatistical data in a GMRF framework using R-inlabru.
    
    \item[3.] Next, a simulation study that reproduces a spatial confounding phenomenon is carried out to assess the coherence of the extensions with the expectations of these methods. Finally, we apply the expanded methods to a case study, linking cadmium (Cd) concentration in terrestrial mosses to Cd concentration in air.
    
    \item[4.] Our findings support the feasibility of our extended approach of spatial confounding resolution methods to geostatistical data using R-INLA in keeping with the previous contexts, although certain precautions and limitations must be considered.
\end{enumerate}
}


Acknowledgements: The authors wish to acknowledge Caroline Meyer, Sébastien Leblond and all the collectors who contributed to the BRAMM survey and those who took care of the samples in the laboratory. The authors would like to thank the staff of the USRAVE laboratory (INRAE-Centre de Bordeaux), accredited by The French Committee for Accreditation, for their analyses of Cd accumulated by mosses. The authors are also thankful to Ilia Ilyin of MSC-East of EMEP for providing the modelled atmospheric deposition and concentration of Cd.\\

Author contributions: Jérémy Lamouroux, Alizée Geffroy, and Isabelle Albert contributed conceptually and formulated the models. Jérémy Lamouroux, Alizée Geffroy provided the practical implementation of the model. All the authors assisted in writing and editing the manuscript.\\

Data availability statement: The code for the simulation study, the models fitting, the terrestrial mosses data, and the EMEP model data are archived on [github].

\baselineskip=\normalbaselineskip

\section{Introduction}\label{sec:1}

\noindent
Spatial confounding has often been studied in recent years to resolve some troubling results in spatial regression models. This issue arises when comparing a regression model with covariates and the same model which includes random spatial effects. \cite{clayton1993spatial} describes the possibility of this phenomenon very early when "the pattern of variation of the covariate is similar to the disease risk, the location may act as a confounder". \cite{gilbert2024causalinferenceframeworkspatial} recently gives four possible causes for spatial confounding: 
"i) Omitted confounder bias: the existence of an unmeasured confounding variable with a spatial structure that influences both the exposure and the outcome;
ii) Regularisation bias: the finite-sample bias of methods that use flexible regression functions like splines or Gaussian processes to control for an unknown function of space;
iii) Random effect collinearity: the change in fixed effect estimates that may come about when spatially-dependent random effects added to a regression model are collinear with the covariates;
iv) Concurvity: the difficulty of assessing the effect of an exposure, which is or is close to a smooth function of space if an arbitrary smooth function of space is also included in a regression model".
The first cause can emerge in any statistical model if it is not in the causal inference framework. The second cause is inherent to statistics which work with finite samples, spatial statistics being perhaps particularly affected as they work with complex modelling such as splines or Gaussian fields. 
The last two causes are particularly interesting to us because they describe a superposed spatial variability. This may be due to spatial collinearity or/and "concurvity", a space-varying co-nonlinearity, between covariates and the outcome variable. It seems to us a modelling challenge to separate this superimposition to make good interpretations and predictions.\

\setlength{\parindent}{30pt} In this article, we explore three methods that have been proposed to address the issue of spatial confounding. One widely used method that interests us is the Restricted Spatial Regression (RSR) proposed by \cite{reich2006effects} and developed later by \cite{HanksEphraim}. This method addresses collinearity between covariates and the spatial random effects introduced into the regression model while preserving the fixed effect estimates obtained without spatial random effects. In addition to RSR, we examined the Spatial+ method proposed by \cite{DupontEmiko} because of its simplicity and flexibility. Spatial+ is a two-step regression method initially developed for models using splines to model spatial random effects. Finally, we explored the recent simplified Spatial+ approach, called here for simplification Spatial+ 2.0, developed by \cite{UrdangarinArantxa}. This modified approach of Spatial+ extends the Spatial+ model on the response by combining the two regression steps into a single regression model. This approach addresses the dependencies between variables from the two distinct steps in Spatial+, aiming to provide more robust estimates. \

\setlength{\parindent}{30pt}These methods have been developed to handle data types such as areal data or geostatistical data using partial thin plate spline models to approximate Gaussian random fields (GRF). Areal data describes situations when space is divided into regions associated with a single aggregated value, such as a mean. This data type is very valuable in epidemiology for disease mapping to highlight regions with high risk and as a first step to discovering potential risk factors related to the response of interest. Many formulations have been proposed to capture unobserved spatial variability as the conditional autoregressive (CAR) prior \citep{besag1974spatial} or the BYM model \citep{besag1991bayesian, riebler2016intuitive}. RSR has been employed much in spatial models for areal data. Other data are geostatistical data, where $Y(s)$ is a random outcome at a specific location and the spatial index $s$ can vary continuously in the fixed subset of $\mathcal{R}^d$. The location $s$ is typically a two-dimensional vector with latitude and longitude but may include altitude \citep{Blangiardo}. This data type is very valuable in spatial modelling for interpolation mapping to highlight geographic variations and as a first step to discovering potential covariates influencing the response of interest. Tobler's First Law of Geography \citep{tobler1970computer} stating that 'everything is related to everything else, but near things are more related than distant things' is fundamental to the spatial approaches of these data. It highlights the spatial autocorrelation often present in geostatistical data, where proximity is critical in the relationships between data points.
In such cases, spline models are a flexible way to capture the space-varying relationships between covariates and the outcome variable in a regression model. Splines are smooth and flexible functions employed to model nonlinear relationships in data, particularly to capture continuous variations in space or time \citep{wahba1990spline}. Spline models have been used to capture continuous spatial variability in the geostatistical data context considered in the Spatial+ method. Its extension, Spatial+ 2.0, is also derived in the thin plate spline context, but its methodology is applied to the (discrete space) ICAR model.\ 

\setlength{\parindent}{30pt} To our knowledge, no study has yet explored applying these three attractive spatial confounding resolution methods to geostatistical data modelled by Gaussian Markov Random Fields (GMRF) in an R-INLA methodology, which is currently the most flexible toolbox for analysing data in continuous space. Thus, the main goal of this work is to propose  extensions of these three existing resolution methods to mitigate spatial confounding from geostatistical data using the R-inlabru package, the extension of the R-INLA package, addressing Bayesian spatial modelling from ecological survey data. 
The INLA (Integrated Nested Laplace Approximation) methodology focuses on models that can be expressed as latent GMRF. Specifying the spatial regression models as Bayesian hierarchical models using latent Gaussian variables, \cite{martinez2013general} proposed a unified formulation for the different specifications of the
spatial dependence between latent variables, all falling under the umbrella of multivariate GMRF. A GMRF is a GRF represented by an undirected graph and defined through its precision matrix (inverse of the covariance matrix). The key advantage of a GMRF is its efficiency, which comes from the sparsity of the precision matrix. The \((i,j)\)-th entry of the precision matrix is nonzero only if nodes $i$ and $j$ are connected in the graph, meaning they are neighbours \citep{rue2005gaussian}. Using this sparsity and nested Laplace approximations for Bayesian inference, R-INLA allows efficient estimations of complex spatial models \citep{lindgren2015bayesian}. It has demonstrated high accuracy for these models while being computationally much faster than Markov Chain Monte Carlo (MCMC) methods \citep{Bakka2018}.
Combined with  the Stochastic Partial Differential
Equation approach  (SPDE) \citep{lindgren2011explicit}, R-INLA can accommodate the spatial continuous space and geostatistical data.

\setlength{\parindent}{30pt} Our interest in such extensions has been motivated by a case study about the potential link between cadmium (Cd) concentrations in territorial mosses collected by the Biosurveillance des Retombées Atmosphériques Métalliques par les Mousses (BRAMM)\footnote{\url{https://bramm.mnhn.fr/}} in France and concentrations of this same element in the air, predicted by a European system (EMEP)\footnote{\url{https://www.emep.int/index.html}}. Several studies \citep{Bates, Ilyin2005modelling, schroder2019moss} show the link between these two concentrations with other datasets. Indeed, mosses' unique morphology, absence of roots, and developed vascular system allow them to function as effective sensors of atmospheric contaminants, as they rely directly on atmospheric deposition for nutrients \citep{Bates}. From our dataset, a linear regression model with a random spatial effect does not find the link but the same model with no spatial random effect estimates a highly significant relationship between the Cd concentration in mosses and the Cd concentration in the air predicted by the EMEP model. 
We have found ourselves in the context of the examples of \cite{reich2006effects} cited by \cite{DupontEmiko} evoking the possibility of spatial confounding. As the data are geostatistical, we have decided to extend the three previous resolution methods in this context using the R-INLA tool, which is currently the most flexible toolbox for analysing data on GRFs. We will fully return to our case study in Section 2.3 of this article.

\setlength{\parindent}{30pt} This paper is structured as follows: Section 2 presents the three approaches we have studied and describes the adaptations developed for their implementation in R-inlabru. We then present a simulation study inspired by \cite{DupontEmiko} to compare these methods for resolving spatial confounding in a GMRF/geostatistical data context (Section 2.2). This study allows us to examine the expected properties of these resolution methods in the proposed extensions. Next, this section describes our case study on Cd concentrations in terrestrial mosses (Section 2.3). Section 3 presents the results obtained from the simulation study and the application. Section 4 discusses all the results, highlighting the precautions and limitations needed to implement these methods on geostatistical  data processed by GMRF using the inlabru package.

\section{Materials And Methods}\label{sec:2} 

\subsection{Three methods to mitigate spatial confounding}

In the following, for the sake of simplicity, the methods and models are presented using a vector $X$, which contains a single covariate corresponding to our simulation and case studies, but note that these methods and models have been developed to handle a matrix $X$ accommodating multiple covariates.
\subsubsection{RSR}

\textbf{Model description}

\noindent It is appropriate to speak of RSR methods in the plural because of the variety of possible approaches \citep{reich2006effects, HodgesReich2010, paciorek2010importance, hughes2013dimension}, but all encapsulated by a general regression \citep{HanksEphraim}: 
\begin{align}
Y(s) &= \beta_{RSR} X(s) + \tilde{U}(s) + \varepsilon(s), \label{eq:RSR}\\
\varepsilon(s)\stackrel{\text{iid}}{\sim} \mathcal{N}(0, \sigma_{\varepsilon}^{2}),\quad
&\tilde{U}(s) = (I - P_{X})U(s),\quad
U(s)\sim\mathcal{N}(0, \Sigma),\quad \nonumber
\end{align}
\noindent where \(Y(s) \) is the outcome variable which $s$ represents the data locations $s = (s_1,..., s_n)$, \(X(s) \) is an observed covariate with unknown fixed effect $\beta_{RSR}$, $\varepsilon(s)$ is an independent and identically distributed errors' noise with residual standard deviation $\sigma_{\varepsilon}$, $I$ the identity matrix, \(P_X = X(X'X)^{-1}X' \) projects onto the space spanned by the columns of \(X = (X(s_1),...,X(s_n))'\), \( U(s)\) is a spatial random effect with mean zero and a covariance matrix $\Sigma$, then \( \tilde{U}(s)\) is a spatial random effect orthogonal to X \citep{HanksEphraim}. Therefore, the method consists of fitting the regression (1) rather than a spatial regression with \({U}(s)\). The idea is to restrict the spatial effects to the space orthogonal to the fixed effect to avoid collinearity between fixed and spatial random effects.\

\setlength{\parindent}{30pt} Two important properties of the RSR model raises \citep{KhanCalder} and \citep{HodgesReich2010}: 
i) the posterior mean of the estimator of $\beta_{RSR}$ is equal to that of the Null model (simple regression model with the covariate and without spatial random effect):
    \begin{equation}
        E[\beta_{RSR} | Y] = E[\beta_{Null} | Y];
        \label{pro:mean}
    \end{equation}
ii) the posterior variance of the estimator of $\beta_{RSR}$ is lower or equal to the posterior variance of the estimator $\beta_{Spatial}$ of the spatial model, including the covariate and the random spatial effect: 
    \begin{equation}
    Var(\beta_{RSR} | Y) \leq Var(\beta_{Spatial} | Y).\label{pro:var}
    \end{equation} 
RSR removes collinearity, and all the spatial variability in the direction of the fixed effect is attributed to the observed covariate.

\setlength{\parindent}{30pt} The RSR model has been mostly applied to areal data. In this context, the intrinsic conditional autoregressive (ICAR) model is commonly used to model the spatial random effect $U(s)$ \citep{HanksEphraim}. Considering an ICAR prior, $U(s)$ is supported on the nodes \(V = 1,2,...,n \text{ of a graph } G = (V,A)\) with $A$ the adjacency matrix where $A_{ij} = 1$ indicates an edge between the nodes $i$ and $j$, while $A_{ij} = 0$ indicates no edge. In this case, the spatial precision matrix (the inverse of the covariance matrix) of $U(s)$ is given by $Q = \Sigma^{-1}$ where $Q = \frac{1}{\sigma^2}(diag(B)-A)$ where $B$ is a matrix with the number of neighbours of each area and $\sigma$ is the standard deviation. Also, RSR has recently been applied to geostatistical data \citep{HanksEphraim}. The spatial random effect, $U(s)$, is modelled using a GRF such as the covariance matrix $\Sigma = \sigma^2 \rho(s,s')$, where $\Sigma$ is the Matèrn covariance function, \( cov(U(s),U(s')) = \sigma^2 \rho(s,s')\), with:
\begin{equation}
    \sigma^2\rho(s,s') = \frac{\sigma^2}{\Gamma(\nu)2^{\nu-1}} \left( \frac{\sqrt{2\nu} d(s,s')}{\rho} \right)^\nu K_\nu \left( \frac{\sqrt{2\nu} d(s,s')}{\rho} \right),
    \label{eq:Matern}
\end{equation}
where $\sigma^2$ is the marginal variance, \(d(s,s')\) is a distance between the points \(s\) and \(s'\), \(\nu\) is the smoothness parameter, $\rho$ is the range parameter, \(\Gamma(\nu)\) is the Gamma function evaluated at \(\nu\), \(K_\nu\) is the modified Bessel function of the second kind \citep{banerjee2003hierarchical}. Hanks et al. \citep{HanksEphraim} constructed an MCMC algorithm to obtain samples from the posterior distributions of the parameters of the RSR model. This approach considers that the constraint  $X'U(s) = 0$ is first sampled from the unconstrained distribution of $U(s)$. \\

\noindent\textbf{Model extension to geostatistical data  using inlabru}\

\noindent The INLA methodology focuses on models that can be expressed as latent GMRF. A spatial regression model  specified as a Bayesian hierarchical model with latent Gaussian variables where the spatial process $U(s)$ is modelled as a GRF with a Matérn covariance function, as defined in (\ref{eq:Matern}), can  be inferred in inlabru package, the recent extension of R-INLA package. A GRF with a Matérn covariance can be expressed as the solution to  Stochastic Partial Differential Equation (SPDE)and an approximate solution to the SPDE is obtained via the finite element method that discretises the spatial domain $\mathbb{R}^d$ into non-intersecting triangles, forming a triangulated mesh; the continuously indexed GRF is then approximated by a discretely indexed GMRF, using finite basis functions defined over this mesh \citep{lindgren2011explicit}. To implement RSR in the context, we must find a $X(s)$ projection onto $U(s)$ to mitigate collinearity between these variables. The idea is to ensure that $\langle X(s), U(s)\rangle=0$, where $\langle,\rangle$ represents the scalar product, enforcing orthogonality and eliminating spatial redundancy. The projection uses $P_X$ in (\ref{eq:RSR}). Implementing this procedure requires specific adjustments to the inference process to avoid concurrently estimating the spatial random and $X$-related effects within the model. The models can be constructed using a formula-like syntax, explicitly defining each latent effect \citep{bachl2019inlabru}. Unlike the syntax typically used in R-INLA, inlabru components provide enhanced functionality, allowing more flexible and complex model specifications, including non-linear predictors. For example, when supplying a formula to R-INLA, the resulting object may refer to the fixed effect $\beta X$, denoted as $X$. This can lead to ambiguity as it becomes unclear whether $X$ refers to the covariate or its associated effect. To address this, R-inlabru provides a shorthand for constructing basic additive models with fixed and random effects, simplifying the implementation of \((I - P_X)U(s)\) as \(U(s) - P_XU(s)\). However, this approach is computationally expensive. To reduce the computational cost, we use the numerical formula:
\[
(I - P_X)U = U - \frac{\sum_{s=s_1}^{s_n}X(s) U(s)}{\sum_{s=s_1}^{s_n}X^2(s)}.
\]
This numerical form avoids matrix operations, significantly improving efficiency while maintaining the projection's key properties. This approach balances computational speed and accuracy, which is crucial for large spatial datasets. 

\setlength{\parindent}{30pt}Additionally, as inference in INLA is inherently Bayesian, it is important to consider the choice of priors for all the model parameters. This requires specifying prior distributions for the parameters that govern the regression coefficient ($\beta_{RSR}$) and the residual standard deviation  parameter ($\sigma_{\varepsilon}$), as well as the   Matérn $U(s)$'s hyperparameters $\rho$ and $\sigma$ (see (4); $\nu$ is fixed to 1, default value in inlabru). Initially, we have opted for relatively vague non-informative priors as defaults. However, during the simulation study, we observed that the priors for $U(s)$ needed to be constrained in order to be in accordance with the  propriety (\ref{pro:mean}) of the RSR model, if not the estimates of the $\beta_{RSR}$ parameter were too much unstable. We then impose the constraints for the penalized complexity (PC) priors \citep{fuglstad2019constructing}: ${P}\mathbb(\rho < \rho_0)$ very close to one and $\mathbb{P}(\sigma > \sigma_0)$ very close to zero for \( U(s)\) to shrink the spatial field towards a base model with $\rho=0$ and $\sigma=0$ rather $\rho=\infty$ and $\sigma=0$ as proposed by default in R-INLA.
\subsubsection{Spatial+}
\textbf{Model description}\\
\noindent The Spatial+ method is a two-step regression approach developed by \cite{DupontEmiko}. 
The first step involves fitting a spatial model to the covariate to eliminate its spatial dependence:
\begin{equation}
    X(s) = f_X(s) + \varepsilon_{X}(s), \quad \varepsilon_X(s) \stackrel{\text{iid}}{\sim} \mathcal{N}(0, \sigma_{X}^{2}),
    \label{eq:step1_spp1}
\end{equation}
where $f_X(s)$  is an unknown bounded function defined on an open bounded domain $\mathbb{R}^d$ which includes the data locations $s = (s_1,..., s_n)$, and $\varepsilon_X(s)$ is an iid errors' noise with residual standard deviation $\sigma_X$. $f_X$ is a smooth function estimated using thin plate splines.

\setlength{\parindent}{30pt} The resulting residual $r^X(s)$, which is the difference between the response covariate $X(s)$ and the estimated spatial residual $\hat{f}_X(s)$,
is introduced in a second regression replacing the covariate in the spatial model:
\begin{align}
       Y(s) &= \beta_{Spatial+} r^X(s) + f(s) + \varepsilon(s), \label{eq:spatial+}\\
       \varepsilon(s)\stackrel{\text{iid}}{\sim}& \mathcal{N}(0, \sigma_{\varepsilon}^{2}),\quad
       r^X(s) = X(s) - \hat{f}_X(s),\nonumber
\end{align}
where $Y(s)$ is the outcome variable, $\beta_{Spatial+}$ is the unknown effect of the resulting residual $r^X(s)$, $f$ is a smooth function to be estimated using thin plate splines, and
$\varepsilon(s)$ is an iid errors' noise with residual standard deviation $\sigma_{\varepsilon}$. Therefore, the method introduces a straightforward modification to the spatial model, replacing the covariate with its residuals after accounting for spatial dependence, which has been eliminated by regressing $X$. Spatial+ preserves the spatial random effect $f(s)$ while addressing collinearity by adjusting the covariate $X(s)$ instead of altering the spatial component of the model. $r^X(s)$ has a reduced or completely absent spatial structure and still contains the covariate information.      

\setlength{\parindent}{30pt}\cite{DupontEmiko} show the good proprieties of the approach to alleviating spatial confounding when traditional spatial models can lead to biased effect estimates due to excessive smoothing of spatial terms, which can confound the effects of covariates. The Spatial+ model addresses this issue by separating the spatial dependence from the covariates, thus avoiding the bias caused by smoothing. It provides unbiased estimates of covariate effects even with spatial smoothing. Notably, in the simulation study, $\hat{\beta}_{Spatial+}$ is, on average, closer to the true $\beta$ value compared to the estimate $\hat{\beta}_{Spatial}$ of the spatial model, including the covariate and the spatial effects.

\setlength{\parindent}{30pt}
As Spatial+ is formulated through the thin plate spline context, the methodology of Spatial+ can be directly applied from geostatistical data as shown by \cite{DupontEmiko}. More recently, a simulation and application case was conducted in the context of areal data, where the spatial random effect was modelled using an ICAR model \citep{UrdangarinArantxa}.
In the geostatistical setting, \cite{DupontEmiko} mention that another approach is possible, instead of applying a thin plate regression spline, using a GMRF model, but they do not implement it.\\ 

\noindent \textbf{Model extension to geostatistical data using inlabru}

\noindent The approach Spatial+ requires two regression steps, it could be a question to do both using inlabru or not. The first one could be done as presented in (5) with \(f_X(s) = f_X(s_1,s_2)\), with \((s_1,s_2)\) as the coordinates (longitude, latitude) of the centroid of the small areas, \(f_X(s_1,s_2) =(f_X(s_{11},s_{12}),...,f_X(s_{n1},s_{n2}))'\) is a smooth function to be estimated using thin plate splines and then the residuals of this regression replace the covariate of interest in  the spatial model as in (6). Also, a more complex Generalized Additive  Model (GAM) could be performed for this regression, as suggested in (5). Then, the regression of interest (6) would be the unique one fitted using inlabru. The use of inlabru requiring Bayesian inference necessitates just one little delicate step to fix parameters' priors, the remainder being a regression model from geostatistical data for which the inlabru package is adapted. The INLA methodology that captures the continuous space by SPDE and finite element method avoids the sensitive choice of the smoothing spline penalties \citep{DupontEmiko}. The choice is also necessary for the first regression, the inlabru option for fitting the first model also seems a good option.

\setlength{\parindent}{30pt}Concerning the priors, relatively vague and non-informative priors can be chosen for the $\beta_{Spatial+}$ parameter and the two residual standard deviations $\sigma_{X}$ and $\sigma$ of the two regression models unless outside information is known as expert knowledge for example. For the hyperparameters of the Matérn GMRFs of the two regression models, we have to take PC priors as implemented in inlabru. As recommended by \citep{fuglstad2019constructing}, a low tail probability of the range parameter, ${P}\mathbb(\rho < \rho_0)=0.01$, and an upper tail probability for the marginal variance parameter, $\mathbb{P}(\sigma > \sigma_0)=0.01$, can be fixed. \cite{fuglstad2019constructing} recommend taking $\sigma_0$ between 2.5 to 40 times the presumed marginal standard deviation, and $\rho_0$ can be set to a value between $\frac{1}{10}$ and $\frac{2}{5}$ of a possible valid range.  These recommended choices were supported by our simulation study and case study (see Sections 2.2 and 2.3 for more details).

\subsubsection{Spatial+ 2.0}
\textbf{Model description}\

\noindent
Recently \cite{UrdangarinArantxa} proposed a simplified version of the Spatial+ approach, which avoids the two regression steps, i.e. it avoids fitting a spatial regression model to the covariate of interest. The alternative to removing the spatial dependence in the covariate is achieved by decomposing the covariate into a linear combination of the eigenvectors of the  precision matrix of the spatial random effect and then performing a unique regression, keeping the short-scale dependence for the regression model with the outcome variable.  
Therefore, the covariate $X(s)$ is decomposed as a linear combination of eigenvectors of $S$ where $S$ is an orthogonal matrix whose columns are the eigenvectors of the spatial precision matrix $Q$ of the spatial random effect ($Q= S\Delta S'$ considering the spectral decomposition of $Q$; $\Delta$ is a $n\times n$ diagonal matrix with the eigenvalues of $Q$ in the main diagonal):
\begin{equation}
X(s) = a_1 S_1 + ... + a_s S_s. 
\label{eq:decomposition}
\end{equation}
Assuming that the collinearity between the fixed and random effects primarily arises from the eigenvectors associated with the lowest non-null eigenvalues, the covariate is split into two parts:
\begin{equation}
X(s) = Z(s) + Z^*(s),\\
\end{equation}
where \(Z^*(s) = a_{s-k}S_{s-k} + ... + a_s S_s\) comprises large-scale eigenvectors associated with the lowest eigenvalues and \(Z(s) = a_1 S_1 +...+a_{s-(k+1)}S_{s-(k+1)}\) represents the remaining eigenvectors. Then, a spatial regression model is fitted on the outcome variable $Y(s)$: 
\begin{equation}
Y(s) = \beta_{Spatial+2.0} Z(s) + U(s) + \varepsilon(s), \label{eq:spatial+ 2.0}\\
\end{equation}
where $Z(s)$ is the spatially decorrelated part of $X$, $U(s) \sim \mathcal{N}(0,\Sigma)$ is the spatial random effect with mean zero and covariance matrix $\Sigma$ and $\varepsilon(s) \sim \mathcal{N}(0, \sigma_{\varepsilon}^2)$ is the iid errors' noise with $\sigma_{\varepsilon}$ as standard deviation. 

\setlength{\parindent}{30pt}\cite{UrdangarinArantxa} show in a simulation study that the Spatial+ 2.0 model  is effective in accurately recovering the true value of the fixed effect regarding the spatial and RSR models. 
However, the choice of the number the number of eigenvectors to keep (or remove) is delicate to have good  $\beta_{Spatial+2.0}$'s estimate proprieties. They recommend removing 7\% and 20\% of the eigenvectors if the spatial covariate pattern is smooth and between 35\% and 43\% if the covariate pattern is less smooth from their simulation and case studies.

\setlength{\parindent}{30pt} 

In \cite{UrdangarinArantxa}, Spatial+ 2.0 is applied to areal data, where model fitting and inference are carried out from a full Bayesian perspective using two main techniques: MCMC and INLA. 
The spatial effect $U(s)$ is modelled using a multivariate intrinsic conditional autoregressive (ICAR) prior with a fixed correlation parameter. Within the Spatial+ framework, $U(s)$ can be simulated by modelling spatial variability as a smooth surface constructed using splines as presented above in Section 2.1.2, however, its extension, called here Spatial+ 2.0, has not yet been applied in a geostatistical data context.\\

\noindent\textbf{Model extension to geostatistical data using inlabru}\

\noindent The Spatial+ 2.0 approach fits a linear model where the covariate $X(s)$ has been previously split into two parts and then uses $k$-selected eigenvectors of the spatial precision matrix $Q$ as regressors.
In the geostatistical context as in the areal data case, selecting the optimal value of $k$ is difficult, especially since, in the continuous spatial case, the dimension of $X$ is generally larger (being the number of measurements over the space). As proposed by \citep{UrdangarinArantxa}, it involves balancing the number of included eigenvectors with minimizing the smallest eigenvalue with the effectiveness of the Bayesian WAIC criterion. 
\setlength{\parindent}{30pt} For the parameter $\beta_{Spatial+ 2.0}$ and the residual standard deviations $\sigma_{\epsilon}$, relatively vague and non-informative priors can be selected. Regarding the hyperparameters of the Matèrn GMRF in the regression model, PC priors must be used as implemented in inlabru. Following the recommendations of \citep{fuglstad2019constructing}, the range parameter can be assigned a low tail probability, with $\mathbb{P}(\rho < \rho_0) = 0.01$ with $\rho_0$ can be chosen between $\frac{1}{10}$ and $\frac{2}{5}$ of an assumed valid range. For the marginal variance parameter, the upper tail probability can be fixed at $\mathbb{P}(\sigma > \sigma_0) = 0.01$ with $\sigma_0$ between 2.5 and 40 times the expected marginal standard deviation. As in the Spatial+ approach, our findings support these priors in the simulation and case studies (see Section 2.2 and 2.3).
\subsection{Simulation study}

In this section, we aim to determine whether the properties previously described for the three methods of addressing confounding effects hold in a simulation scenario where geostatistical data are generated and fitted with the inlabru package. In this scenario, the true relationship between the covariate and the response variable ($\beta$) is known, and the random effect $U(s)$ is assumed to follow a GRF with a Matérn covariance matrix. We compare the models based on the accuracy of $\beta$'s estimates and the global Bayesian model selection criteria WAIC \citep{gelman2014understanding}.

\subsubsection{Data}
We generate 50 independent data, constructing each with an observed covariate \(\mathbf{x} = (x(1), \dots, x(n))'\) and response data \(\mathbf{y} = (y(1), \dots, y(n))'\) such as:
\begin{align*}
    x(s) &= 0.35 z_x(s) + \varepsilon_x(s),\\
    y(s) &= \beta x(s) + u(s) + \varepsilon_y(s),
\end{align*}
where $s=1,...,n$ with $n=500$ randomly selected locations within a continuous spatial field \([0,10] \times [0,10]\), \(\mathbf{z_x} = (z_{x}(1), \dots, z_{x}(n))'\) is the spatial residual of the covariate  collected at the $n=500$ locations with a Matérn short-scale spatial correlation, $\epsilon_x(s) \sim \mathcal{N}(0,\sigma_x^2 = 0.1)$, $\beta$ = 3 is the true covariate effect, \(u(s) = z_u(s) - z_x(s)\) are the spatial effects where \(\mathbf{z_u} = (z_{u}(1), \dots, z_{u}(n))'\) is the spatial random effect collected at the $n=500$ locations with a Matérn large-scale spatial correlation, $\epsilon_y(s) \sim \mathcal{N}(0,\sigma_y = 1)$. This approach to creating spatial confounding from geostatistical data is similar to \cite{DupontEmiko}'s one in their simulation study except that here the continuous spatial random fields have Matérn spatial covariance matrix created from an R-INLA approach, i.e. creating for each field a triangulated mesh which is the support of the finite element method's solution, a GMRF approximative solution, of the SPDE coupled with the parameters of the Matérn covariance function. We employ two meshes of 1,283 nodes, the ratio between the number of observations and the mesh's size being 0.39. We see with other mesh sizes that this ratio is a good trade-off between the accuracy of the GMRF representation and the computational cost. We put $\theta_1 = 2$ and $\theta_2 = 0.4$ for $\mathbf{z_x}$, and $\theta_1 = 0$ and $\theta_2 = 1$ for $\mathbf{z_u}$ in R-INLA  ($\theta_1 = log(\sigma) - log(\sigma_0)$ and $\theta_2 = log(\rho) - log(\rho_0)$) to control our two GRFs (vague PC priors have been fixed: $\mathcal{P}(\rho < 0.5) = 0.5$ and $\mathcal{P}(\sigma > 0.5) = 0.5$).
\subsubsection{Models}
We fit five regression models to each of the 50 datasets comprised of the response variable $Y$ and the covariate of interest $X$ using the inlabru package. They are:\\
1. A linear regression model (Null model) with no spatial effect given by: 
    \begin{equation}
       Y(s) = \beta_{Null} X(s) + \varepsilon(s), \label{eq:null} \quad \varepsilon(s) \stackrel{\text{iid}}{\sim} \mathcal{N}(0, \sigma_{\varepsilon}^{2}),
   \end{equation}
where $\beta_{Null}$ is the unknown fixed effect to estimate and $\varepsilon(s)$ the iid errors’ noise with the standard deviation $\sigma_{\varepsilon}$. Vague and non-informative priors are fixed for the $\beta_{Null}$'s parameter and the residual standard deviation $\sigma_{\varepsilon}$.\\
2. A spatial model with a spatial random effect given by:
\begin{align}
       Y(s) &= \beta_{Spatial} X(s) + U(s) + \varepsilon(s), \label{eq:spatial}\\
        \varepsilon(s) \stackrel{\text{iid}}{\sim} \mathcal{N}(0, &\sigma_{e}^{2}),\quad
        U(s) \sim \mathcal{N}(0, \sigma_{u}^2 \rho(s, s')), \nonumber
\end{align}
where $\beta_{Spatial}$ is the covariate fixed effect to estimate, $\varepsilon(s)$ is an iid errors' noise with the standard deviation $\sigma_{\varepsilon}$, $U(s)$ is a spatial random effect modelled as a GRF with mean zero and Matérn covariance matrix. Vague and non-informative priors are given to the parameter $\beta_{Spatial}$ and the residual standard deviation $\sigma_{\varepsilon}$. PC priors are taken as: $\mathbb{P}(\rho < 0.05) = 0.05$ and $\mathbb{P}(\sigma > 3) = 0.05$.\\
3. A RSR model given in (\ref{eq:RSR}) with $\Sigma =  \sigma^2 \rho(s, s')$ given in (\ref{eq:Matern}), where $\nu$ has been fixed to 1. Vague and non-informative priors are given to the parameter $\beta_{RSR}$ and the residual standard deviation $\sigma_{\varepsilon}$. PC priors are taken as $\mathbb{P}(\rho < 15) = 0.9999$ and $\mathbb{P}(\sigma > 1.5) = 0.0001$ as mentionned in Section 2.1.1, to shrink the model towards a baseline model with $\rho=0$ and $\sigma=0$.\\
4. A Spatial+ model given in (\ref{eq:step1_spp1}) and (\ref{eq:spatial+}), with $f_X(s)$ and $f(s)$ $\sim \mathcal{N}(0, \Sigma)$, with $\Sigma$ given in (\ref{eq:Matern}) with $\nu = 1$. Vague and non-informative priors are given to the parameter $\beta_{Spatial+}$ and the residual standard deviations $\sigma_{X}$ and $\sigma_{\varepsilon}$. PC priors, for the first regression, are taken as: $\mathbb{P}(\rho^X < 0.01) = 0.01$ and $\mathbb{P}(\sigma^X > 0.15) = 0.01$ and, for the second regression, are taken as: $\mathbb{P}(\rho^Y < 0.05) = 0.05$ and $\mathbb{P}(\sigma^Y > 3) = 0.05$.\\
5. A Spatial+ 2.0 given by (\ref{eq:spatial+ 2.0})  with $\Sigma =  \sigma^2 \rho(s, s')$ given in (\ref{eq:Matern}), where $\nu$ has been fixed to 1. Vague and non-informative priors are given to the parameter $\beta_{Spatial+2.0}$ and the residual standard deviation $\sigma_{\varepsilon}$. PC priors are taken as $\mathbb{P}(\rho < 0.05) = 0.05$ and $\mathbb{P}(\sigma > 3) = 0.05$.
We retain 400 of the 500 available eigenvectors based on minimising the WAIC criterion, calculated for each of the 500 possible models. This procedure has been replicated 50 times for each dataset, consistently converging around 400 eigenvectors.


\subsection{Case study: Application on terrestrial mosses} 
\subsubsection{Data}
\begin{figure}
    \centering
    \begin{subfigure}[t]{0.49\textwidth}
        \centering
        \includegraphics[width=\linewidth]{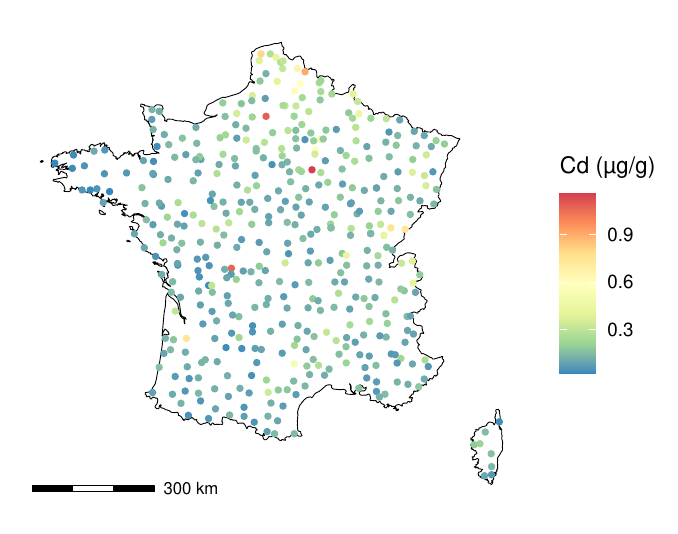}
    \end{subfigure}\hfill
    \begin{subfigure}[t]{0.5\textwidth}
        \centering
        \includegraphics[width=\linewidth]{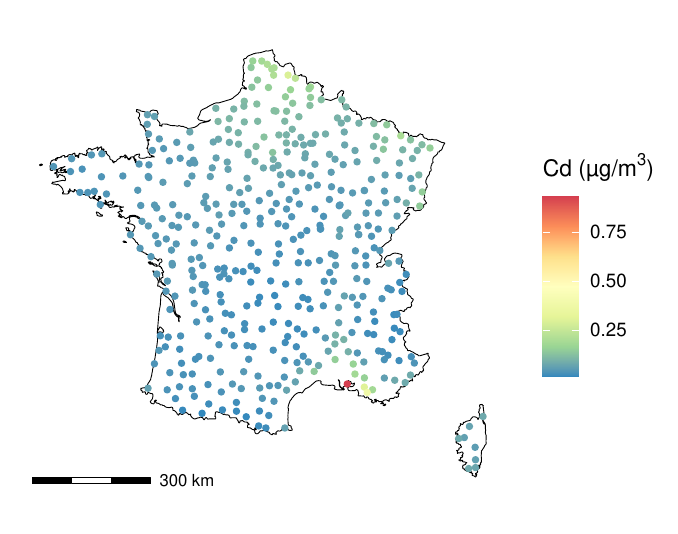}
    \end{subfigure}
    \caption{Concentration maps of Cd across France over 445 sites (BRAMM sites) in 2016: Mosses' Cd concentration measurements in $\mu g / g$ (left); Air Cd concentrations in $\mu g / m^3$ from the physical model EMEP values (right).}
    \label{fig:Cd_mosse_EMEP}
\end{figure}

We apply the proposed extensions in a spatial confusing case we have encountered studying the link between Cd concentrations measured in terrestrial mosses throughout France and Cd concentration predicted in air in France by an EMEP\footnote{\url{https://www.emep.int/index.html}} model, an atmospheric transport model used to predict metal concentrations across Europe. 
We analyse a mosses dataset collected in 2016 as part of the BRAMM\footnote{\url{https://bramm.mnhn.fr/}} scheme, which covered 445 sites across France \citep{leblond2011dispositif}. 
At each of the 445 sites, $s$, a sample of a mosses' mixture of a unique species over $s$ ($4km^2$) is collected, and its Cd concentration, $Y(s)$  after that, is measured in the lab. Using a linear regression model, we relate this concentration to an air Cd concentration value obtained at location $s$, $X(s)$ after that, from the values predicted by the EMEP model for 2016 on a 0.1º x 0.1º longitude-latitude grid. To obtain the value at the mosses' location, $X(s)$, we used the nearest EMEP value from the site $s$. Figure \ref{fig:Cd_mosse_EMEP} shows the values $Y(s)$ and $X(s)$ at the $s$ location.

We initially fitted a Null model, and the parameter $\beta_{Null}$ was highly positively significant ($p<5.07e^{-7}$) but a residual spatial autocorrelation of the response values was still there (indicating from Moran's index z-scores, $p<2.2e^{-16}$). We then fitted a spatial model to the dataset, and the parameter $\beta_{spatial}$ was no longer significant. These findings have questioned us, and then we turned to models that could resolve potential spatial confounding. In the next section, we provide details of the fitted model using inlabru. 
\subsubsection{Models}
1. A Null model as defined in (\ref{eq:null}) with a fixed intercept $\beta_{0Null}$, where $Y(s)$, $s \in (1,...,445)$, are a logarithmic transformation of Cd mosses' concentrations to reduce the skewness of their distribution, $X(s)$, are the Cd concentrations in air calculated from the EMEP's values as described above. Vague and non-informative priors are fixed for the parameters $\beta_{0Null}$ and $\beta_{Null}$, and the residual standard deviations $\sigma_{\varepsilon}$.\\
2. A spatial model as defined in (\ref{eq:spatial}) with a fixed intercept $\beta_{0Spatial}$, with ($Y(s)$, $X(s)$) defined as in the Null model above. Vague and non-informative priors are given to the parameters $\beta_{0Spatial}$ and $\beta_{Spatial}$ and the residual standard deviations $\sigma_{\varepsilon}$. PC priors are taken as: $\mathbb{P}(\rho < 6e^{4}) = 0.05$ and $\mathbb{P}(\sigma > 5) = 0.05$.\\
3. A RSR model given by (\ref{eq:RSR}) with a fixed intercept $\beta_{0RSR}$, $\Sigma =  \sigma^2 \rho(s, s')$ given in (\ref{eq:Matern}), where $\nu$ has been fixed to 1, with ($Y(s)$, $X(s)$) defined as in the Null model above. Vague and non-informative priors are given to the parameters  $\beta_{0RSR}$ and $\beta_{RSR}$, and the residual standard deviations $\sigma_{\varepsilon}$. PC priors are taken as $\mathbb{P}(\rho < 18e^{6}) = 0.9999$ and $\mathbb{P}(\sigma > 2.5) = 0.0001$. Note that in the projection $P_X$, $X$ includes the ones of the intercept.\\
4. A Spatial+ model given by (\ref{eq:step1_spp1}) and (\ref{eq:spatial+}), with a fixed intercept $\beta_{0Spatial+}$ in the second regression model,  $f_X(s)$ and $f(s)$ $\sim \mathcal{N}(0, \Sigma)$, with $\Sigma$ given in (\ref{eq:Matern}) with $\nu = 1$, with ($Y(s)$, $X(s)$) defined as in the Null model above. Vague and non-informative priors are given to the  $\beta_{0Spatial+}$ and $\beta_{Spatial+}$ parameters, and the residual standard deviations $\sigma_{X}$ and $\sigma_{\varepsilon}$. PC priors for the first regression are taken as: $\mathbb{P}(\rho^X < 13e^{4}) = 0.05$ and $\mathbb{P}(\sigma^X > 13) = 0.05$ and for the second regression are taken as: $\mathbb{P}(\rho^Y < 6e^{4}) = 0.05$ and $\mathbb{P}(\sigma^Y > 5) = 0.05$.\\
5. A Spatial+ 2.0 model given by (\ref{eq:spatial+ 2.0})  with a fixed intercept $\beta_{0Spatial+2.0}$,  $\Sigma =  \sigma^2 \rho(s, s')$ given in (\ref{eq:Matern}), where $\nu$ has been fixed to 1, with ($Y(s)$, $X(s)$) defined as in the Null model above. Vague and non-informative priors are given to the $\beta_{0Spatial+2.0}$ and $\beta_{Spatial+2.0}$ parameters, and the residual standard deviations $\sigma_{\varepsilon}$. PC priors are taken as $\mathbb{P}(\rho < 6e^{4}) = 0.05$ and $\mathbb{P}(\sigma > 5) = 0.05$.
We retain 310 of the 445 available eigenvectors (76.63\%) based on minimising the WAIC criterion.

\section{Results}\label{sec:3}
\subsection{Simulation study}

\begin{figure}[H]
    \centering
    \begin{subfigure}[t]{0.48\textwidth}
        \centering
        \includegraphics[width=\linewidth]{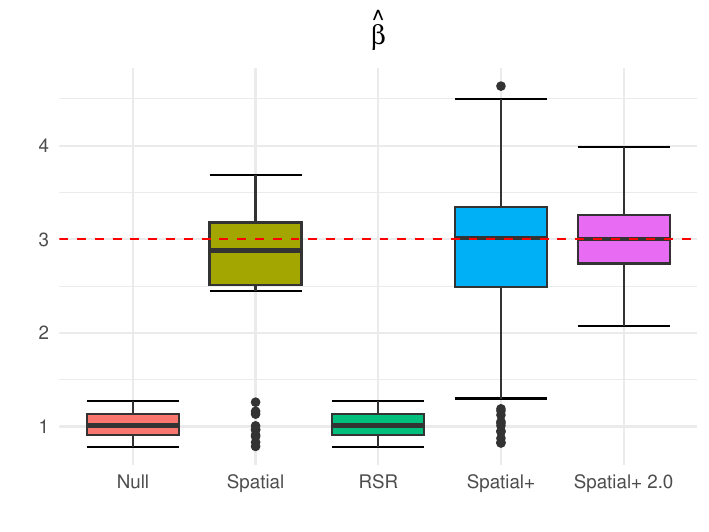}
        \label{fig:beta}
    \end{subfigure}\hfill
    \begin{subfigure}[t]{0.48\textwidth}
        \centering
        \includegraphics[width=\linewidth]{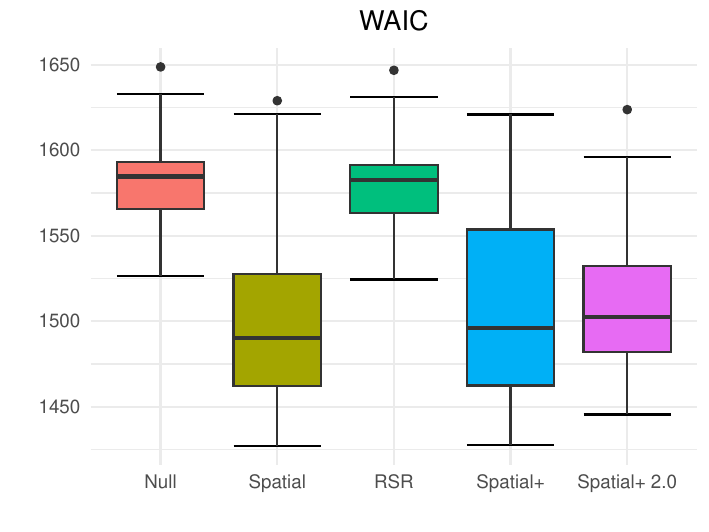}
        \label{fig:WAIC}
    \end{subfigure}
    \caption{Simulation results: estimated covariate effect $\hat{\beta}$ (posterior mean) for each model and each data replicate. The true covariate effect is $\beta=3$ (red dashed line) (left); WAIC values for each model and each data replicate (right).} 
    \label{fig:beta_WAIC}
\end{figure}

Figure \ref{fig:beta_WAIC} gives the box-plots of the posterior means of the estimated $\hat{\beta}$ for the five models fitted to the 50 data replicates and box-plots of their WAIC values. For each model, Table \ref{tab:Table1} summarises the mean of the posterior means, $\hat{\beta}$, the empirical standard deviations of the posterior mean, ESD, the mean estimated standard errors, $\overline{SE}$, the mean of the DIC values, the mean of the WAIC values, and the empirical coverage of 95\% credibility intervals, CI, over the 50 replicated data.


\begin{table}[h!]
\centering
\caption{Table \ref{tab:Table1} summarises the mean of the posterior means, $\hat{\beta}$, the empirical standard deviations of the posterior mean, ESD, the mean estimated standard errors, $\overline{SE}$, the mean of the DIC values, the mean of the WAIC values, and the empirical coverage of 95\% credibility intervals, CI, over the 50 replicated data, for each model.}
\begin{tabular}{lccccccc}
\toprule
Model      & $\hat{\beta}$   & ESD &  $\overline{SE}$ &  DIC   & WAIC  & CI  \\
\midrule
Null        &   1.031  & 0.136 & 0.149 &   1582.51     &   1582.529    &     0   \\
Spatial     &   2.649  & 0.800  & 0.303 &  1498.956    &   1500.482    &     76    \\
RSR         &   1.031  & 0.136  & 0.149 &   1580.524    &   1580.525    &     0     \\
Spatial+    &   2.781  & 1.063  &  0.303 & 1501.736    &   1503.327    &     74    \\
Spatial+2   &   2.999   & 0.375 &  0.412 & 1503.197    &   1507.389    &     96       \\
\bottomrule
\end{tabular}
\label{tab:Table1}
\end{table}

If one first examines the Null and spatial models, one observes very different results as expected due to the superposed variability created in each dataset between the covariate and the spatial random effect (spatial confounding). Here, one observes that the spatial model is closer to the true value with coverage at 76\% even if the estimates are in a large range (ESD = 0.8 and see Figure \ref{fig:beta_WAIC} (left)).


\setlength{\parindent}{30pt} In this context of geostatistical data and R-INLA methodology, the RSR model preserves the theoretical properties outlined in (\ref{pro:mean}) and (\ref{pro:var}): all the posterior means of  ${\beta}_{RSR}$ and ${\beta}_{null}$ are identical (see Figure \ref{fig:beta_WAIC} (left) and Table \ref{tab:Table1}) and the RSR model has got small estimated standard errors, ($\overline{SE}$), than the spatial model indicating reduced estimated variance for ${\beta}_{RSR}$ compared to ${\beta}_{Spatial}$. RSR remains overly restrictive for addressing spatial confounding, even if its WAIC values are a slightly lower than those of the Null model. 
The Spatial+ models, which involve two regression steps in inlabru, produce a mean $\hat{\beta}$ estimate of 2.781, with an empirical coverage of 74\%. 
One observed in Figure \ref{fig:beta_WAIC} (left) that the estimates' range are the largest (ESD = 1.063 in Table 1).
The Spatial+ 2.0 models, where 400 eigenvectors have been kept as explained in Section 2.2.2, 
yield a mean estimated $\hat{\beta}$ of 2.999, with an empirical coverage of 96\%, 
and the range of its estimates are narrow the true $\beta = 3$ (Figure \ref{fig:beta_WAIC} (left), ESD = 0.375 in Table \ref{tab:Table1}). \
\setlength{\parindent}{30pt}Comparing all the DIC and WAIC values, the Spatial, Spatial+ and Spatial+ 2.0 models give almost similar results. Our findings are following the expected properties of the three methods. Note nevertheless that the PC priors of the RSR models have been taken as  $\mathbb{P}(\rho < \rho_0)$ very close to one and $\mathbb{P}(\sigma > \sigma_0)$ very close to zero for \({U}(s)\) to shrink the model towards a baseline model with $\rho=0$ and $\sigma=0$.

\subsection{Case study}
Table \ref{tab:Table2} gives the posterior mean of the parameter $\beta$, the 95\% CI, DIC and WAIC for each model fitted using inlabru. As mentioned, the covariate effect is significant with the Null model and no longer with the spatial model. The RSR model has results very close to those of the Null model. Spatial+ results are very close to the Spatial model. The Spatial+2.0 model has a higher estimate value for the covariate effect but is still non-significant. DIC and WAIC criteria give similar results, preferring Spatial, Spatial+ and Spatial+2.0 models, the Spatial+2.0 model having the smallest value. 
\begin{table}[h!]
\centering
\caption{The posterior mean of the covariate effect $\beta$, quantiles at 2.5\% and 97.5\% of $\beta$, DIC value, WAIC value, and estimated standard error of the posterior mean (SE) for each model fitted in inlabru for the case study.}
\begin{tabular}{lccccccc}
\toprule
Model      & $\hat{\beta}$ & 0.025quant  & 0.975quant   & DIC   & WAIC  & SE       \\
\midrule
Null        &    2.306     &      1.422 & 3.191    &  769.07     &     775.49    & 0.451  \\
Spatial     &     0.230    &     -0.820& 1.280     &  628.45     &     629.28    & 0.536   \\ 
RSR         &    2.302     &      1.415 & 3.189    & 773.21      &     778.21    & 0.451    \\
Spatial+    &    0.213     &     -0.837 & 1.258    & 628.84      &     629.57    & 0.535     \\
Spatial+2.0   &    0.748     &     -0.845 & 2.341    &  628.41     &     628.25  & 0.813      \\
\bottomrule
\end{tabular}
\label{tab:Table2}
\end{table}\

\begin{figure}[ht]
    \centering
    \includegraphics[width=0.9\textwidth]{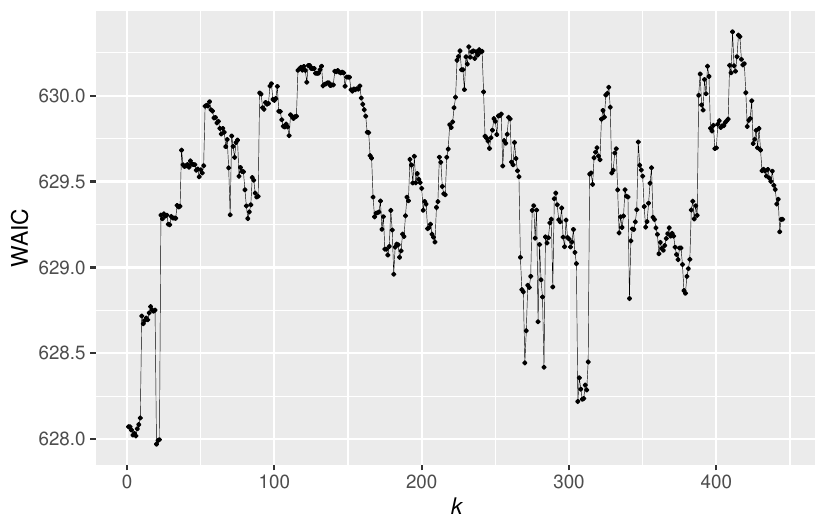}
    \caption{WAIC value of the Spatial+ 2.0 models versus the number $k$ of eigenvectors kept in the model.}
    \label{fig:plot_MSE_sp2_445}
\end{figure}

For the Spatial+ 2.0 model,
we have selected a subset of eigenvectors based on the WAIC values of the 445 different models, splitting $X(s)$ into two parts. Figure \ref{fig:plot_MSE_sp2_445} gives these different values for all models. The model with the smallest WAIC is chosen, i.e., 310 eigenvectors kept (the smallest WAIC values with low eigenvectors $[0, 20]$ kept in the models have been left out as suggested by \cite{urdangarin2023evaluating} because these models could be too conservative inflating parameters standard errors). 

\begin{figure}[H]
    \centering
    \begin{subfigure}[t]{0.49\textwidth}
        \centering
        \includegraphics[width=\linewidth, trim = {0 2.5cm 0 1.5cm}, clip]{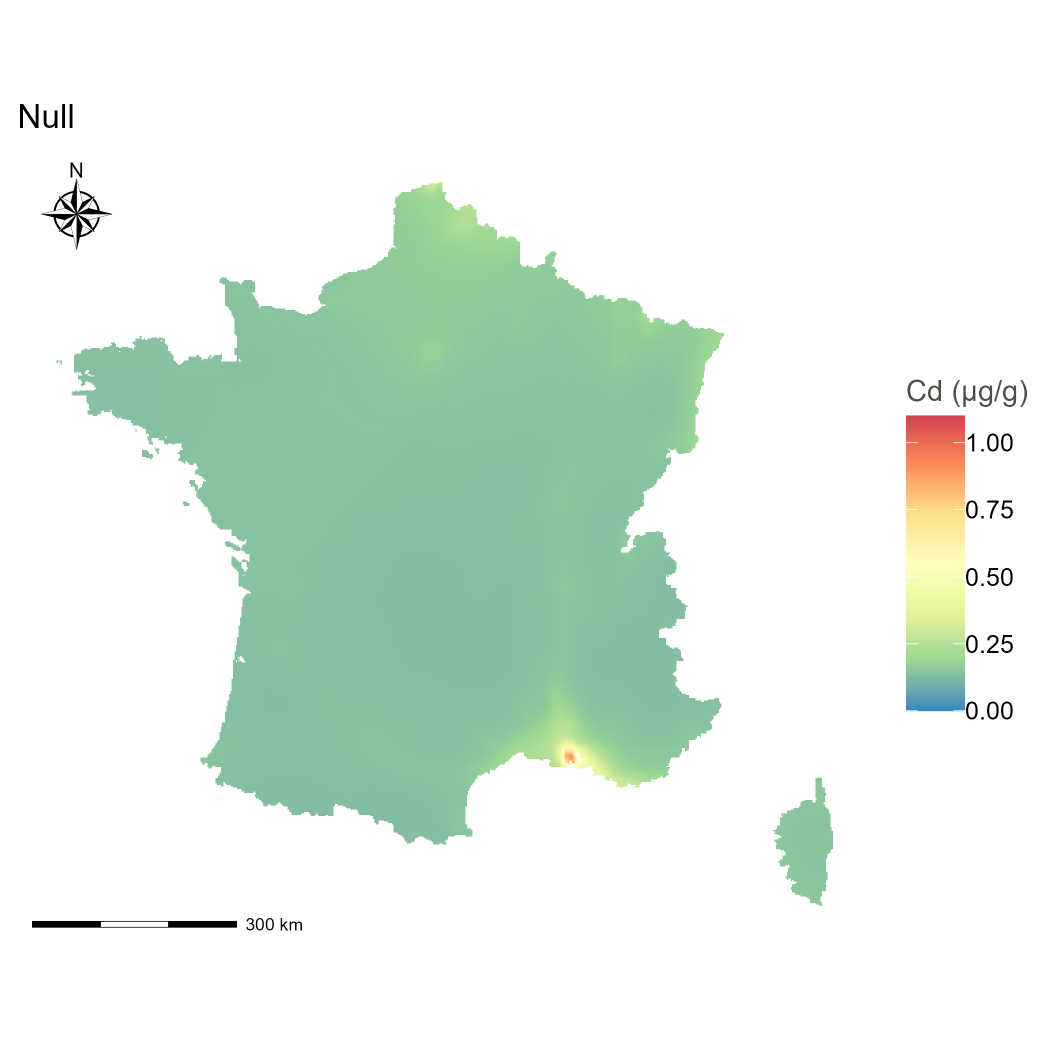}
    \end{subfigure}\hfill
    \begin{subfigure}[t]{0.49\textwidth}
        \centering
        \includegraphics[width=\linewidth, trim = {0 2.5cm 0 1.5cm}, clip]{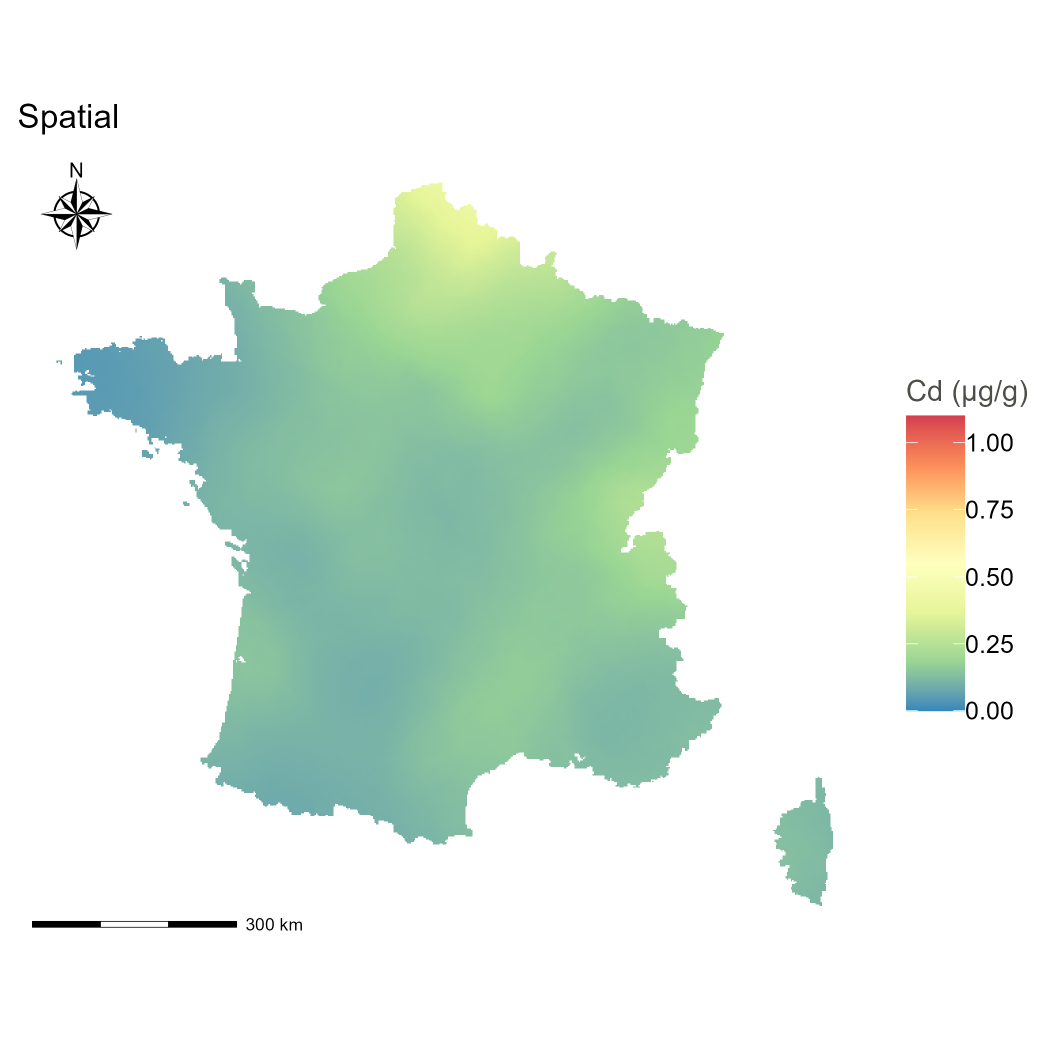}
    \end{subfigure}
        \begin{subfigure}[t]{0.49\textwidth}
        \centering
        \includegraphics[width=\linewidth]{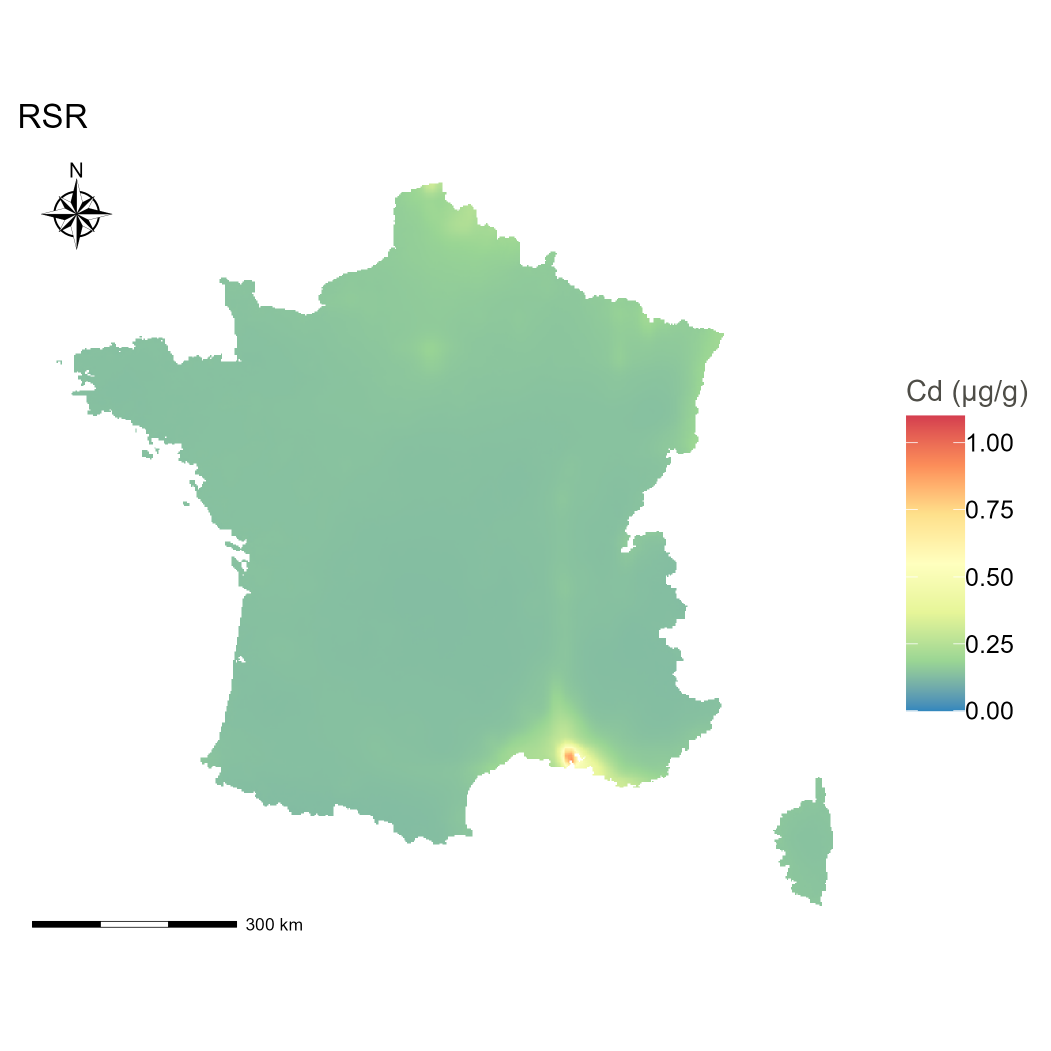}
    \end{subfigure}\hfill
        \begin{subfigure}[t]{0.49\textwidth}
        \centering
        \includegraphics[width=\linewidth]{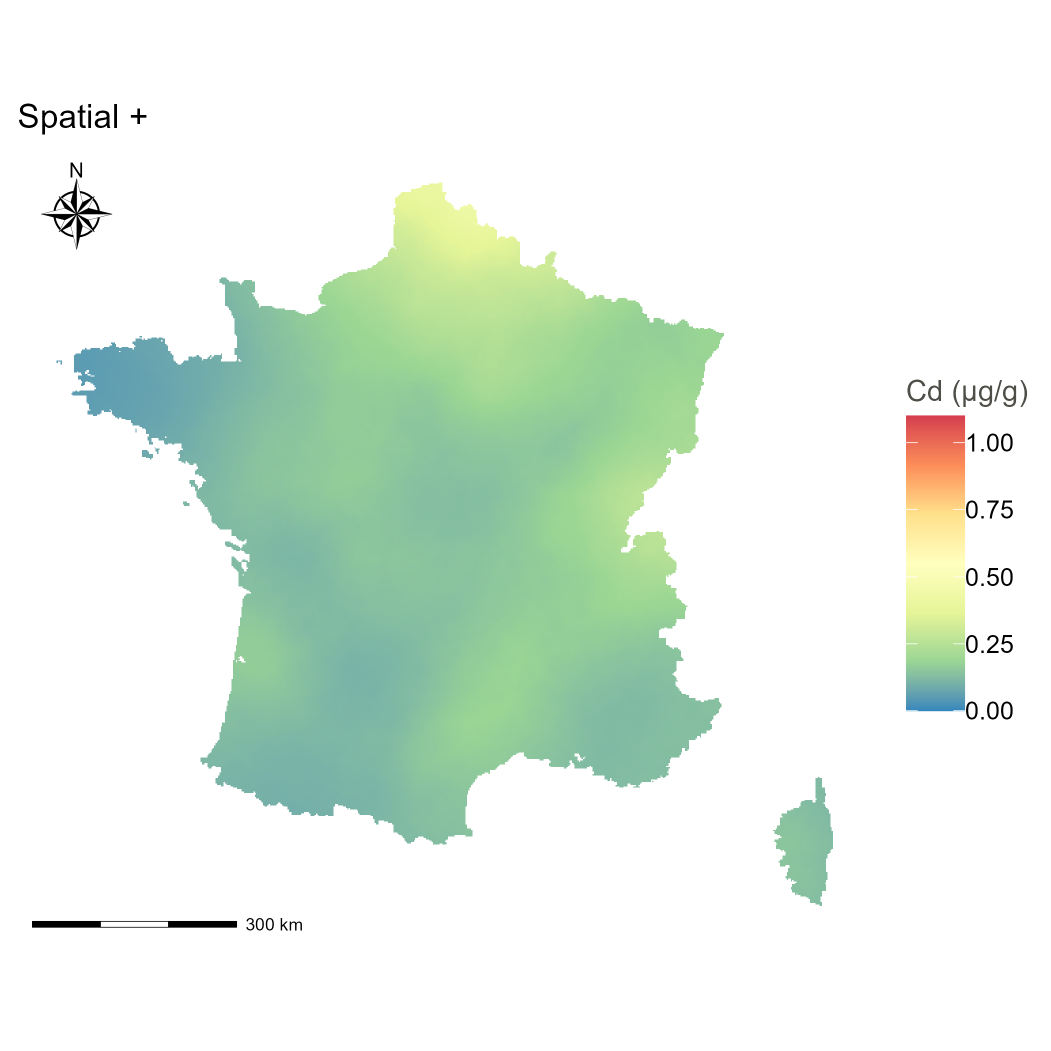}
    \end{subfigure}
    \label{map:spatial+}
    \caption{Cd concentration mosses' prediction maps with a common scale between 0 and 1.1 µg/g: Null model (Average Cd prediction values are in the range [0.12 ; 0.98] (µg/g)); spatial model (Average Cd prediction values are in the range [0.04 ; 0.41] (µg/g)); RSR model ( Average Cd prediction values are in the range [0.12 ; 1.1] (µg/g)); Spatial+ model (Average Cd prediction values are in the range [0.04 ; 0.39] (µg/g)).}
    \label{map:map_pred}
\end{figure}

Figure \ref{map:map_pred} displays the posterior medians prediction maps for Cd concentrations in mosses, generated using the Null, Spatial, RSR, and Spatial+ models. A grid consisting of 139,673 pixels of 4$km^2$ was constructed to cover the map of France. The Spatial+2.0 method necessitates  the spectral  decomposition of the spatial precision matrix $Q$, a 139,673 $\times$ 139,673 matrix, this decomposition can not be obtained numerically. So, we can not produce the prediction maps with the method at such a resolution. We tried a grid of 30,000 pixels, but it was still too large.  Decreasing the map resolution again seems unreasonable to us.
As expected from the model estimates, the Null and RSR models produce identical prediction maps, while the Spatial and Spatial+ maps are very similar. The latter two maps more accurately highlight a northeastern zone with relatively high values, consistent with observed data. In contrast, the Null and RSR maps predict elevated values around the city of Marseille in the southeast of France, which are found in the $X(s)$ values (see Figure \ref{fig:Cd_mosse_EMEP} (right)). By incorporating a larger spatial structure, the Spatial and Spatial+ models capture more spatial correlation (Figure \ref{map:map_pred} (left)). 

\section{Discussion}\label{sec:4}
This paper explores how recent methods (RSR, Spatial+, Spatial+ 2.0) that try to mitigate spatial confounding can be implemented in the context of geostatistical data modelled by latent GMRF in an R-INLA methodology. We assess their implementation using inlabru, the recent R-package which extends R-INLA for modelling ecological data. We have generated simulated datasets in a geostatistical data context of spatial confounding to compare the results of five models: a Null model (without spatial modelling), a spatial model, an RSR model, a Spatial+, and Spatial+ 2.0 model. Null and spatial models give sharply contrasting results, as expected. 
The Spatial+ 2.0 has the best empirical coverage at 95\% and the more closely estimate of the fixed effect $\beta$.
But our simulation framework is very close to that of \cite{DupontEmiko}, so this result may not be generalised to any spatial confounding scheme. The RSR model gives results close to the Null model, as it attributes all spatial variability to the covariate. Regarding the simulation study, we have to say that the results were fairly dependent on the PC priors on the spatial random effects, showing how difficult it is to separate spatial patterns. Note, however, that the PC prior to the range parameter for the spatial random effect of the RSR method has to be set to shrink the model to a baseline model with a range of zero. Otherwise, the estimation of the effect of the covariate was highly unstable (results not shown).\

\setlength{\parindent}{30pt}We are interested in a real case study of terrestrial mosses to explore the relationship between Cd concentration in mosses and air Cd concentration given by the EMEP physical model. This subject can be an important issue to have reliable proxy sources of air pollution for population-based epidemiological studies investigating the health effects of individual metals or groups of metals \citep{lequy2023long}. In this context, we have explored the three methods cited above, but there is no way of finding the effect of the covariate; however, Spatial+2.0 increases the effect of the covariate (but does not become significant). Perhaps an eigenvector separation based on expert opinion (giving the spatial scale of the possible effect of the covariate) would be a better separation criterion than a WAIC criterion. We have asked our bryologist colleagues about this, but they think it is difficult to define a priori on this scale, especially because of localised effects that could change the effect of the covariate on the territory. It would, therefore, be interesting to consider the inclusion of other covariates in the model, such as the wind speed and/or the tree canopy on the sampling site. We already explored some of these possibilities in \cite{Lamouroux2024}.\

\setlength{\parindent}{30pt}Other methods for dealing with spatial confounding could also be considered, such as gSEM \citep{thaden2018structural}, TGMRF \citep{prates2015transformed} and others. They would then need to be implemented, and their extensions must be validated in the R-INLA methodology. Another approach is, of course, to make more complex spatial modelling, such as random-effects GAMs or SVC (spatial varying coefficient) models, which would allow the effect of the covariate to be non-fixed over the territory, both bringing flexibility to the model and tracks for the inclusion of other explanatory covariates.

\setlength{\parindent}{30pt}In conclusion, mitigating spatial confounding is a critical challenge for spatial statistics, because when covariates and response have close spatial patterns, it is not easy to distinguish what belongs to the covariates from what does not. Unlike longitudinal or cluster data, there are no replications (intra-individual or intra-cluster) in general to distinguish spatial patterns. Here, everything is determined by single data at location $s$. It would seem, then, that the answer lies partly with the expert, who knows his/her field and can choose between models. Ultimately, this is the case with all statistical modelling work since statistics merely provide more or less probable interpretations.

\newpage

\bibliographystyle{chicago} 

\end{document}